\documentclass[aps,pre,twocolumn,superscriptaddress,notitlepage]{revtex4-2}
\usepackage{amsmath}
\usepackage{amsfonts}
\usepackage{amssymb}
\usepackage{lmodern,dsfont}
\usepackage{graphicx}
\usepackage[usenames,dvipsnames]{xcolor}
\usepackage{bm}
\usepackage{xr}
\usepackage[english=american]{csquotes}

\newcommand{\av}[1]{\langle #1 \rangle}
\newcommand{\Av}[1]{\left\langle #1 \right\rangle}
\newcommand{\nn}{\nonumber \\}
\newcommand{\n}{\nonumber}

\newcommand{\grad}{\bm{\nabla}}

\renewcommand{\eqref}[1]{Eq.~(\ref{#1})}

\begin{document}

\author{Andreas Dechant}
\affiliation{Department of Physics \#1, Graduate School of Science, Kyoto University, Kyoto 606-8502, Japan}
\title{Thermodynamic constraints on the power spectral density in and out of equilibrium}
\date{\today}

\begin{abstract}
The power spectral density of an observable quantifies the amount of fluctuations at a given frequency and can reveal the influence of different timescales on the observable's dynamics.
Here, we show that the spectral density in a continuous-time Markov process can be both lower and upper bounded by an expression involving two constants that depend on the observable and the properties of the system.
In equilibrium, we identify these constants with the low- and high-frequency limit of the spectral density, respectively; thus, the spectrum at arbitrary frequency is bounded by the short- and long-time behavior of the observable.
Out of equilibrium, on the other hand, the constants can no longer be identified with the limiting behavior of the spectrum, allowing for peaks that correspond to oscillations in the dynamics.
We show that the height of these peaks is related to dissipation, allowing to infer the degree to which the system is out of equilibrium from the measured spectrum.
\end{abstract}

\maketitle

In most realistic physical systems, the observed dynamics is generally the result of processes on a multitude of timescales.
One way to reveal the contributions of different timescales is to analyze the dynamics in the frequency-domain, taking the Fourier transform of a measured time-series.
The average magnitude of the resulting quantity is called power spectral density (PSD) \cite{Zie06,Yat14}.
Intuitively, the spectrum quantifies how much the signal varies at a given frequency, allowing to isolate the contributions from different timescales.

The Wiener-Khinchin theorem \cite{Wie30,Khi34} provides a one-to-one relation between the PSD and the autocorrelation of a stationary process.
In the PSD, oscillations in the time-dependent signal appear as peaks, whose position and width are directly related to the frequency and decay rate of the oscillations, respectively.
As a consequence, the PSD is one of the most fundamental tools for analyzing and visualizing the properties of measured signals. 
It has found widespread applications in physics \cite{Mol69,Bon98,Ber04,Kra19}, climate science \cite{Web01} and medicine \cite{Kij87}, to only name a few, and its estimation from measured data is well-developed \cite{Bin67,Mar01,Geo03}.

However, despite being widely used in data analysis, little is known about constraints on the PSD and its relation thermodynamic properties of the system.
This is in contrast to fluctuations of observables in the time-domain, for which universal bounds in terms various thermodynamic quantities have recently been derived in the literature.
For example, the thermodynamic uncertainty relation \cite{Bar15,Gin16,Hor20} relates the fluctuations of a current to entropy production, allowing to estimate dissipation from the measured fluctuations of a current and other observables \cite{Li19,Dec21c}.
More recently, the statistics of transition times have been used to infer both dissipation and the topological structure \cite{Ski21,Mee22,Har22} of the system's configuration space.

\begin{figure}
\includegraphics[width=.47\textwidth]{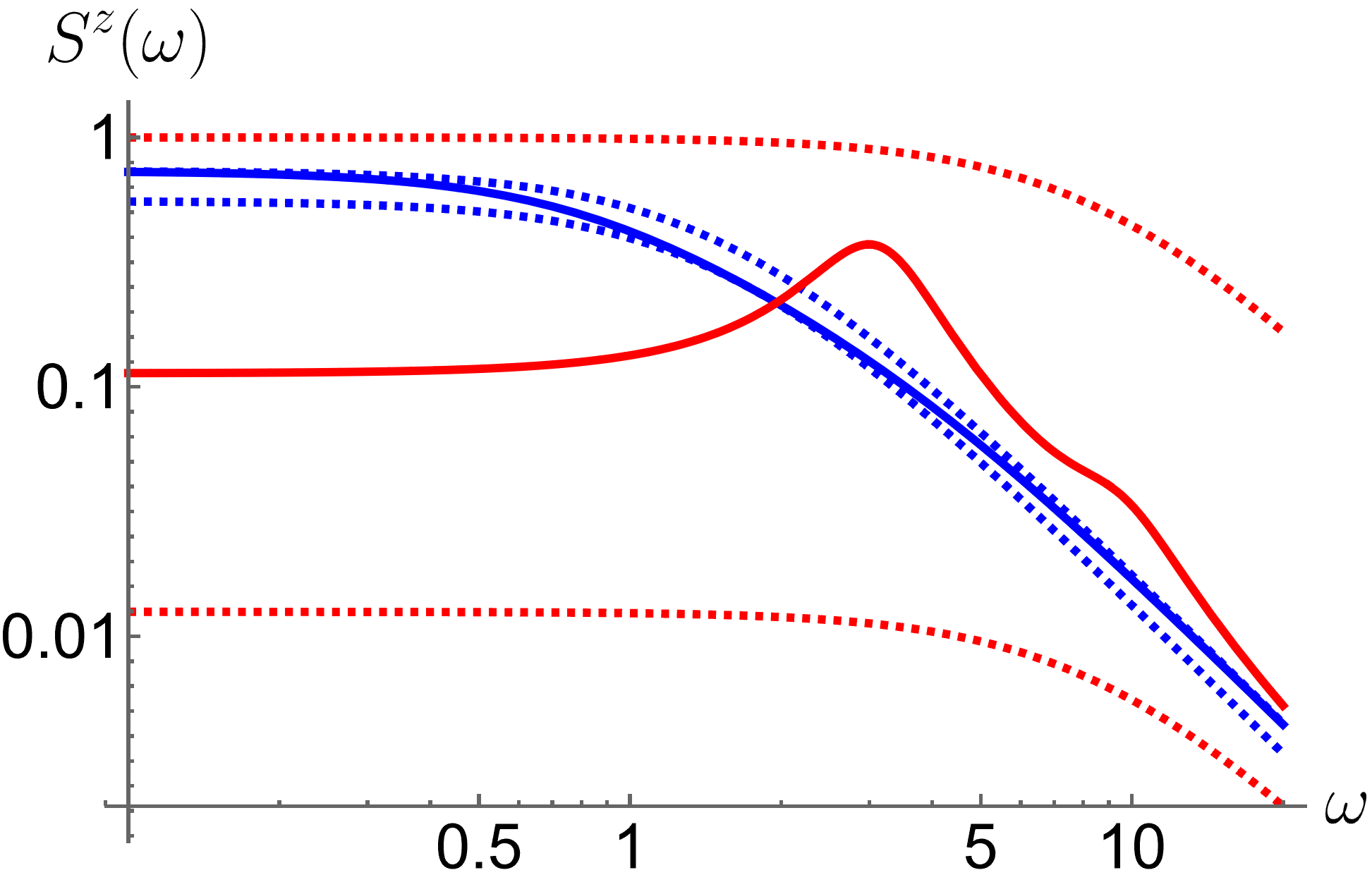}
\caption{Illustration of the lower and upper bounds on the PSD. 
In equilibrium (blue), the PSD is a monotonously decaying function of $\omega$ and bounded from below and above by its asymptotic limits (blue dashed, \eqref{spectral-bound-lower-upper} with \eqref{c-equilibrium}).
Out of equilibrium (red), the behavior of the PSD is more complicated and generally exhibits peaks that correspond to oscillations in the system. 
In this case the lower and upper bound (red dashed, \eqref{spectral-bound-lower-upper-noneq}) depend on the amount of dissipation in the system.}
\label{fig-sd-bound}
\end{figure}
Given these results and the practical importance of the PSD, it is natural to wonder whether the latter obeys similar bounds, and whether they allow us to obtain thermodynamic quantities from the measured spectrum.
The first main result of this Letter are the lower and upper bounds for the PSD $S^z(\omega)$ of an observable $z$ in the steady state of overdamped Langevin and Markov jump processes,
\begin{align}
\frac{1}{C^\infty + \omega^2 C^0} \leq \frac{S^z(\omega)}{2\text{Var}_\text{st}(z)} \leq \frac{1}{\frac{1}{C^0} + \frac{\omega^2}{C^\infty}} \label{spectral-bound-lower-upper} .
\end{align}
Here, $\text{Var}_\text{st}(z)$ denotes the steady-state fluctuations of the observable.
If the system is in equilibrium, the constants $C^0$ and $C^\infty$ correspond to the zero- and infinite-frequency limit of the PSD, respectively,
\begin{align}
C_\text{eq}^0 = \frac{S_\text{eq}^Z(0)}{2 \text{Var}_\text{st}(z)}, \quad C_\text{eq}^\infty = \lim_{\omega \rightarrow \infty} \bigg( \frac{\omega^2 S^z_\text{eq}(\omega)}{2 \text{Var}_\text{st}(z)} \bigg) \label{c-equilibrium} .
\end{align}
For equilibrium systems, where the PSD is a monotonously decaying function of $\omega$, \eqref{spectral-bound-lower-upper} provides a relation between the spectrum at intermediate frequencies and its asymptotic behavior.
As we demonstrate below, this information can be used to check the consistency of extrapolated data from measurements over a finite frequency range.

The bound \eqref{spectral-bound-lower-upper} also holds for out-of-equilibrium systems, however, the identification of the constants $C^0$ and $C^\infty$ with the limiting behavior of the PSD no longer holds.
While $C^0$ and $C^\infty$ are generally expressed in terms of variational expressions (see below for details), we can derive the upper bounds,
\begin{subequations}
\begin{align}
C^0 &\leq \frac{1}{\lambda^*} \\
C^\infty &\leq C^\infty_\text{diss} = \lim_{\omega \rightarrow \infty} \bigg( \frac{\omega^2 S^z(\omega)}{2 \text{Var}_\text{st}(z)} \bigg) + \frac{\Delta z^2 \sigma_\text{st}}{4 \text{Var}_\text{st}(z)}  .
\end{align} \label{c-bounds}%
\end{subequations}
Here, $\Delta z = z_\text{max} - z_\text{min}$ denotes the range of the observable.
$\lambda^*$ is the spectral gap, defined as the first non-zero eigenvalue of the symmetrized generator of the dynamics, which does not distinguish between equilibrium and non-equilibrium systems.
By contrast, a non-zero steady-state rate of entropy production $\sigma_\text{st}$ is a clear indication that the system is out of equilibrium.
Inserting the bounds on $C^0$ and $C^\infty$ into \eqref{spectral-bound-lower-upper} results in the looser but more explicit bounds on the spectral density
\begin{align}
\frac{1}{C^\infty_\text{diss} + \frac{\omega^2}{\lambda^*}} \leq \frac{S^z(\omega)}{2\text{Var}_\text{st}(z)} \leq \frac{1}{\lambda^* + \frac{\omega^2}{C^\infty_\text{diss}}} \label{spectral-bound-lower-upper-noneq} ,
\end{align}
which are our second main result.
Since $C^\infty_\text{diss}$ differs from its equilibrium value by a term proportional to the entropy production, this implies that the degree to which the PSD can deviate from the monotonic decay of an equilibrium system is controlled by the amount of dissipation.
The bounds \eqref{spectral-bound-lower-upper} and \eqref{spectral-bound-lower-upper-noneq} are illustrated in Fig.~\ref{fig-sd-bound}.
The concrete system from which the data in Fig.~\ref{fig-sd-bound} is obtained is discussed in  Appendix S~IV.


\begin{figure}
\includegraphics[width=.47\textwidth]{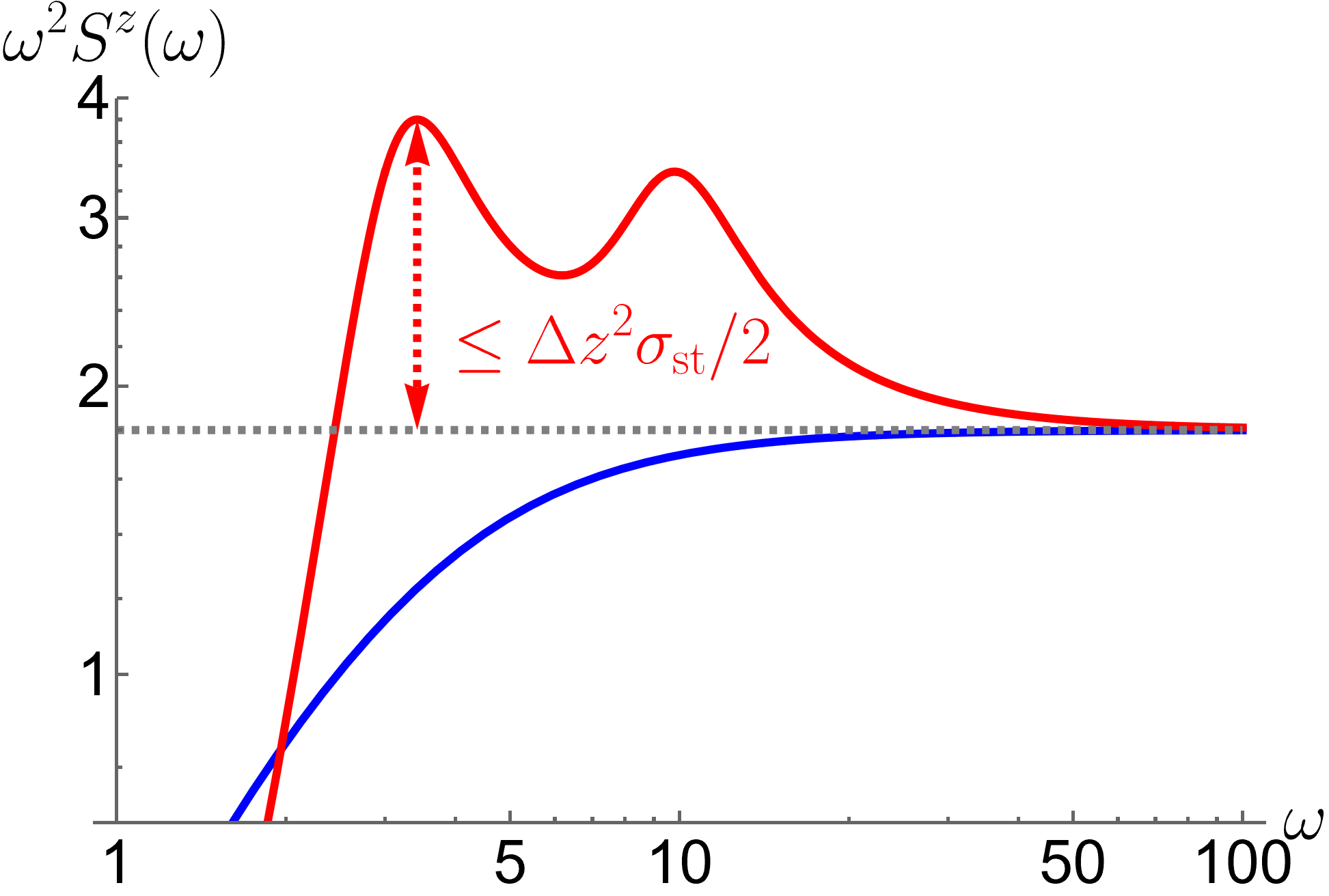}
\caption{Relation between peaks in the PSD and entropy production. 
In equilibrium (blue), the PSD always approaches its high-frequency limit (dashed gray) from below. 
Out of equilibrium (red), by contrast, the high-frequency asymptote may be exceeded by the intermediate-frequency PSD; the difference between the latter and the former gives a lower bound on the steady state entropy production rate, see \eqref{spectral-tur-omega}.}
\label{fig-entropy-estimate}
\end{figure}
We can further use \eqref{spectral-bound-lower-upper-noneq} to relate the measured PSD to the entropy production rate,
\begin{align}
\sigma_\text{st} \geq \frac{2}{\Delta z^2} \bigg( \omega^2 S^z(\omega) - \lim_{\omega \rightarrow 0} \big( \omega^2 S^z(\omega) \big) \bigg) \label{spectral-tur-omega}.
\end{align}
This lower bound, which constitutes our third main result, involves the finite-frequency PSD, and allows estimating the entropy production rate directly from a measurement of the former:
Plotting $\omega^2 S^z(\omega)$ as a function of $\omega$, the resulting graph converges to a constant value at large frequencies, which characterizes the short-time fluctuations of the observable.
Equilibrium processes cannot exhibit oscillations and therefore the PSD cannot have any local maxima, approaching its high-frequency limit from below.
For non-equilibrium processes, however, oscillations are possible and appear as local maxima in the PSD, whose height is related to the amount of dissipation by \eqref{spectral-tur-omega}, see Fig.~\ref{fig-entropy-estimate}.
We remark that, despite this intuitive relation, the quantity $\omega^2 S^z(\omega)$ can have a local maximum even when $S^z(\omega)$ does not, and therefore \eqref{spectral-tur-omega} can give a positive estimate even if there are no oscillations in the system.

\textit{Diffusion processes and PSD.}
In the following, we restrict the discussion to diffusion processes with continuous configuration space $\bm{x}(t) \in \mathbb{R}^d$, however, we stress that the bounds \eqref{spectral-bound-lower-upper}, \eqref{spectral-bound-lower-upper-noneq} and \eqref{spectral-tur-omega} also apply to jump processes on a discrete state space.
The time-evolution of the configuration is described by the overdamped Langevin equation
\begin{align}
\dot{\bm{x}}(t) = \bm{f}(\bm{x}(t)) + \bm{G} \cdot \tilde{\bm{\xi}}(t) \label{langevin} .
\end{align}
Here, $\bm{f}(\bm{x}) \in \mathbb{R}^d$ is a drift vector that describes the systematic forces acting in the system, while $\bm{G} \in \mathbb{R}^{d \times d}$ is a matrix of rank $d$ describing the coupling to the $d$ mutually independent Gaussian white noises $\tilde{\bm{\xi}}(t)$.
In the following, we will focus on the time-independent steady state with probability density $p_\text{st}(\bm{x})$, which is attained in the long time limit under mild assumptions.
It is the solution of the steady-state Fokker-Planck equation \cite{Ris86}
\begin{align}
0 &= - \grad \big( \bm{\nu}_\text{st}(\bm{x}) p_\text{st}(\bm{x}) \big) \qquad \text{with} \label{fpe-steady} \\
 \bm{\nu}_\text{st}(\bm{x}) &= \bm{f}(\bm{x}) - \bm{B} \grad \ln p_\text{st}(\bm{x}) \n ,
\end{align}
where $\bm{B} = \bm{G} \bm{G}^\text{T}/2$ is the positive definite diffusion matrix and $\bm{\nu}_\text{st}(\bm{x})$ is called the local mean velocity.
We can distinguish two fundamentally different types of dynamics:
If $\bm{f}(\bm{x}) = -\bm{B} \grad \psi(\bm{x})$ with some scalar potential $\psi(\bm{x})$, then the steady-state solution is the Boltzmann-Gibbs density $p_\text{st}(\bm{x}) \propto e^{-\psi(\bm{x})}$ and the local mean velocity vanishes, corresponding to an equilibrium dynamics satisfying detailed balance \cite{Ris86}.
On the other hand, if $\bm{f}(\bm{x})$ cannot be written as the gradient of a potential, then we have $\bm{\nu}_\text{st}(\bm{x}) \neq 0$ and a non-equilibrium steady state with entropy production rate
\begin{align}
\sigma_\text{st} = \Av{\bm{\nu}_\text{st} \bm{B}^{-1} \bm{\nu}_\text{st}}_\text{st},
\end{align}
where $\av{\ldots}_\text{st}$ denotes an average with respect to the steady state density.
We now measure an observable $z(\bm{x}(t))$ whose value depends on the configuration $\bm{x}(t)$, and thus fluctuates due to the random evolution of the latter.
We define the finite-time Fourier transform of the time-series of $z$ over the measurement interval $[0,\tau]$,
\begin{align}
\hat{z}_{\tau}(\omega) = \int_0^\tau dt \ e^{i \omega t} z(\bm{x}(t)) .
\end{align}
The PSD of $z$ is defined as the long-time limit of the variance,
\begin{align}
S^z(\omega) = \lim_{\tau \rightarrow \infty} \bigg( \frac{1}{\tau} \Big( \Av{ \big\vert \hat{z}_\tau(\omega) \big\vert^2} - \big\vert \Av{\hat{z}_\tau(\omega)} \big\vert^2 \Big) \bigg) \label{spectral-definition}.
\end{align}
By the Wiener-Khinchin theorem \cite{Wie30,Khi34}, this is equal to the Fourier transform of the covariance of $z(\bm{x})$,
\begin{align}
S^z(\omega) = \int_{-\infty}^\infty dt \ e^{i \omega t} \text{Cov}(z(t), z(0)) .
\end{align}
Let us recall a few well-known results for the PSD.
First, by definition, its zero-frequency component is equal to the long-time limit of the variance of the time-average $\bar{z}_\tau = \int_0^\tau dt \ z(\bm{x}(t))/\tau$,
\begin{align}
S^z(0) = \lim_{\tau \rightarrow \infty} \Big( \tau \text{Var}(\bar{z}_\tau) \Big) \equiv D^{\bar{z}} \label{low-frequency-limit} .
\end{align}
The high-frequency limit, on the other hand, is related to the short-time fluctuations of the displacement $dz = z(\bm{x}(t+dt)) - z(\bm{x}(t))$,
\begin{align}
\lim_{\omega \rightarrow \infty} \big( \omega^2 S^z(\omega) \big) = \lim_{dt \rightarrow 0} \bigg( \frac{\text{Var}(dz)}{dt} \bigg) \equiv D^{dz}_0 \label{high-frequency-limit} .
\end{align}
The latter quantity can be computed explicitly \cite{Dec23},
\begin{align}
D^{dz}_0 = 2 \Av{\grad z \bm{B} \grad z}_\text{st}.
\end{align}
These relations make explicit the intuitive notion that low frequencies capture the long-time fluctuations of the dynamics, whereas high frequencies describe the short-time fluctuations.

\textit{Variational expression for the PSD and bounds.}
The bound \eqref{spectral-bound-lower-upper} can be derived from a variational expression for the PSD, whose proof is provided in Appendix S~I.
There, we show the identity
\begin{widetext}
\begin{align}
S^z(\omega) &= 2 \sup_{\chi \in \mathbb{C}} \inf_{\eta \in \mathbb{C}} \Bigg[ 2 \Re \bigg(\text{Cov}_\text{st}\big( z + \grad \eta \bm{\nu}_\text{st} - i \omega \eta , \chi^* \big)  \bigg)  \label{spectral-variational}+ \Av{\grad \eta^* \bm{B}\grad \eta}_\text{st} - \Av{\grad \chi^* \bm{B} \grad \chi}_\text{st}  \Bigg] .
\end{align}
\end{widetext}
In this expression, the maximization and minimization are performed over complex valued, configuration-dependent functions $\chi(\bm{x})$ and $\eta(\bm{x})$.
$\text{Cov}_\text{st}$ denotes the covariance with respect to the steady state and $\Re$ the real part.
While \eqref{spectral-variational} is difficult to evaluate explicitly, it serves as a useful starting point for deriving bounds on the PSD: Restricting the domain of or choosing a specific function for $\eta(\bm{x})$ yields an upper bound, while doing the same for $\chi(\bm{x})$ results in a lower bound.
We first choose $\eta(\bm{x}) = i z(\bm{x})$.
Then, after some simplifications (see the Appendix S~II for details), we can formulate the upper bound
\begin{align}
\frac{S^Z(\omega)}{2 \text{Var}_\text{st}(z)} \leq \frac{1}{\frac{1}{C^0} + \frac{\omega^2}{C^\infty}} \label{spectral-upper}
\end{align}
which reproduces the upper bound in \eqref{spectral-bound-lower-upper}.
The lower bound, on the other hand, is obtained by choosing $\chi(\bm{x}) = z(\bm{x})$ and simplifying, resulting in
\begin{align}
&\frac{S^Z(\omega)}{2 \text{Var}_\text{st}(z)} \geq \frac{1}{C^\infty + \omega^2 C^{0}} \label{spectral-lower} .
\end{align}
The general expressions for $C^0$ and $C^\infty$ are found as
\begin{subequations}
\begin{align}
C^0 &= \sup_{\chi \in \mathbb{R}} \bigg[ \frac{\text{Cov}_\text{st}(z,\chi)^2}{\text{Var}_\text{st}(z) \Av{\grad \chi \bm{B} \grad \chi}_\text{st}} \bigg], \label{c-0} \\
C^{\infty} &= \sup_{\chi \in \mathbb{R}} \Bigg[ \frac{\Av{\grad z \bm{B} \grad z}_\text{st} +  \frac{\Av{\chi \grad z \bm{\nu}_\text{st}}_\text{st}^2}{\Av{\grad \chi \bm{B} \grad \chi}_\text{st}}}{\text{Var}_\text{st}(z)} \Bigg] \label{c-infty}.
\end{align} \label{c-expressions}%
\end{subequations}
This proves \eqref{spectral-bound-lower-upper}, that is, there exist two constants that yield a lower and an upper bound on the PSD at any frequency.
Directly comparing the lower and upper bound, we also find the relation $C^0 C^\infty \geq 1$ between the two constants.
Note that $C^0$ and $C^\infty$ are generally expressed in terms of variational expressions depending on the steady-state probability density, the local mean velocity and the diffusion matrix.

In order to obtain more explicit bounds, we first remark that using any upper bound on $C^0$ and/or $C^\infty$ also results in weaker, lower and upper bounds on the PSD.
For $C^0$, an upper bound can be obtained by using the covariance inequality, $\text{Cov}_\text{st}(z,\chi')^2 \leq \text{Var}_\text{st}(z) \text{Var}_\text{st}(\chi')$, which yields
\begin{align}
C^0 \leq \sup_{\chi \in \mathbb{R}} \bigg[ \frac{\text{Var}_\text{st}(\chi)}{\Av{\grad \chi \bm{B} \grad \chi}_\text{st}} \bigg] = \frac{1}{\lambda^*} .
\end{align}
The latter expression is equivalent to a well-known variational formula for the first non-zero eigenvalue of an equilibrium system (see, e.~g., chapter 6.6.2 in Ref.~\cite{Ris86})
\begin{align}
\lambda^* = \inf_{\chi \in \mathbb{R}} \bigg[ \frac{\Av{\grad \chi \bm{B} \grad \chi}_\text{st}}{\text{Var}_\text{st}(\chi)} \bigg] \label{spectral-gap}.
\end{align}
In the case of \eqref{langevin}, this corresponds to the dynamics in the conservative force field $\bm{f}(\bm{x}) = \bm{B} \grad \ln p_\text{st}(\bm{x})$, which has the same steady state as \eqref{langevin} but a vanishing local mean velocity.
Since $\lambda^*$ is obtained from a symmetrized version of \eqref{langevin}, it is referred to as the spectral gap of the system.
We note that this quantity has recently appeared in bounds on fluctuations and speed limits \cite{Bak22,Bao23}, confirming its fundamental role in characterizing the dynamics.
On the other hand, an upper bound on the second term in $C^\infty$ can be obtained as follows,
\begin{align}
\frac{\Av{\chi \grad z \bm{\nu}_\text{st}}_\text{st}^2}{\Av{\grad \chi \bm{B} \grad \chi}_\text{st}} &= \frac{\Av{(z-z_0) \grad \chi \bm{\nu}_\text{st}}_\text{st}^2}{\Av{\grad \chi \bm{B} \grad \chi}_\text{st}} \\
&\leq \Av{(z-z_0)^2 \bm{\nu}_\text{st} \bm{B}^{-1} \bm{\nu}_\text{st}}_\text{st} \n .
\end{align}
In the first step, we added an arbitrary constant $z_0$ to $z(\bm{x})$ and integrated by parts using \eqref{fpe-steady}; in the second step, we used the Cauchy-Schwarz inequality.
Minimizing the right-hand side with respect to $z_0$, this can be written as
\begin{align}
\frac{\Av{\chi \grad z \bm{\nu}_\text{st}}_\text{st}^2}{\Av{\grad \chi \bm{B} \grad \chi}_\text{st}} \leq \sigma_\text{st} \text{Var}_\sigma(z) ,
\end{align}
where we defined the entropy-rescaled probability density
\begin{align}
p_\sigma(\bm{x}) = \frac{\bm{\nu}_\text{st} \bm{B}^{-1} \bm{\nu}_\text{st}}{\sigma_\text{st}} p_\text{st}(\bm{x})
\end{align}
and $\text{Var}_\sigma$ denotes the corresponding variance.
If the observable is bounded, $z_\text{min} \leq z(\bm{x}) \leq z_\text{max}$ with $\Delta z = z_\text{max} - z_\text{min}$, then we have the upper bound
\begin{align}
\text{Var}_\sigma(z) \leq \frac{\Delta z^2}{4} .
\end{align}
We then obtain an upper bound on $C^{\infty}$,
\begin{align}
C^{\infty} \leq \lim_{\omega \rightarrow \infty} \bigg( \frac{\omega^2 S^z(\omega)}{2 \text{Var}_\text{st}(z)} \bigg) + \frac{\Delta z^2 \sigma_\text{st}}{4 \text{Var}_\text{st}(z)} ,
\end{align}
where we used \eqref{high-frequency-limit}, proving \eqref{c-bounds}.

To establish a relation between the PSD and the entropy production rate, we use the trivial upper bound $C^0 \leq \infty$ and find from \eqref{spectral-upper}, 
\begin{align}
\frac{S^Z(\omega)}{2 \text{Var}_\text{st}(z)} \leq \frac{C^\infty}{\omega^2} .
\end{align}
Using the above upper bound on $C^\infty$ and rearranging yields \eqref{spectral-tur-omega}.
We remark that a non-trivial lower bound on the entropy production rate is only obtained for bounded observables.
In practice, this is not a serious constraint, since from any unbounded observable, we easily obtain a bounded one by truncating the measured values of $z(\bm{x}(t))$.
In that case, the truncation values enter the bound as additional optimization parameters.

\textit{Equilibrium and non-equilibrium systems.}
Let us now specialize the above results to equilibrium systems ($\bm{\nu}_\text{st}(\bm{x}) = 0$).
As we show in Appendix S III, in this case, the variational expression \eqref{spectral-variational} reduces to an optimization with respect to real functions,
\begin{align}
S^z_\text{eq}(\omega) = 2 \sup_{\chi \in \mathbb{R}} \inf_{\eta  \in \mathbb{R}} \Bigg[ \frac{\text{Cov}_\text{st}(z,\chi)^2}{\Av{\grad \chi \bm{B} \grad \chi}_\text{st} + \omega^2 \frac{\text{Cov}_\text{st}(\chi,\eta)^2}{\Av{\grad \eta \bm{B} \grad \eta}_\text{st}}} \Bigg] \label{spectral-eq-variational} .
\end{align}
This describes a monotonously decaying function of $\omega$, reflecting the monotonous decay of correlations in equilibrium \cite{Tu08}.
Setting $\omega = 0$, we obtain the identity
\begin{align}
\frac{S^z_\text{eq}(0)}{2 \text{Var}_\text{st}(z)} = C^0 .
\end{align}
The corresponding relation for $C^\infty$ in \eqref{c-equilibrium} follows directly form \eqref{high-frequency-limit} and \eqref{c-infty}, setting $\bm{\nu}_\text{st}(\bm{x}) = 0$.
Since $C^0$ also provides a global upper bound on $S^z(\omega)$, both in and out of equilibrium, the non-equilibrium PSD at any frequency is upper bounded by the zero-frequency component in the equilibrium system with the same steady state.
In particular, we have $S^z(0) \leq S^z_\text{eq}(0)$, that is, the zero-frequency component of a non-equilibrium system is reduced compared to equilibrium.
Since the total power,
\begin{align}
\int_{-\infty}^{\infty} d\omega \ S^z(\omega) = 2 \pi \text{Var}_\text{st}(z),
\end{align}
is the same in both cases, this implies that, generally, the effect of driving the system out of equilibrium is shifting power from lower to higher frequencies, see Fig.~\ref{fig-sd-bound} for illustration.
\eqref{spectral-tur-omega} relates the enhanced power at intermediate frequencies to the rate of entropy production.
Setting $\omega = 0$ in the lower bound in \eqref{spectral-bound-lower-upper-noneq}, we can also obtain a relation between entropy production and the reduction in the zero-frequency component,
\begin{align}
\sigma_\text{st} \geq \frac{2}{\Delta z^2} \bigg( \frac{4 \text{Var}_\text{st}(z)^2}{S^z(0)} - \lim_{\omega \rightarrow \infty} \big(\omega^2 S^z(\omega) \big) \bigg) \label{spectral-tur-0} .
\end{align}
This relation is equivalent to the bound on the fluctuations of the time-average of $z(\bm{x})$ derived in Ref.~\cite{Dec23}.

\begin{figure}
\includegraphics[width=0.47\textwidth]{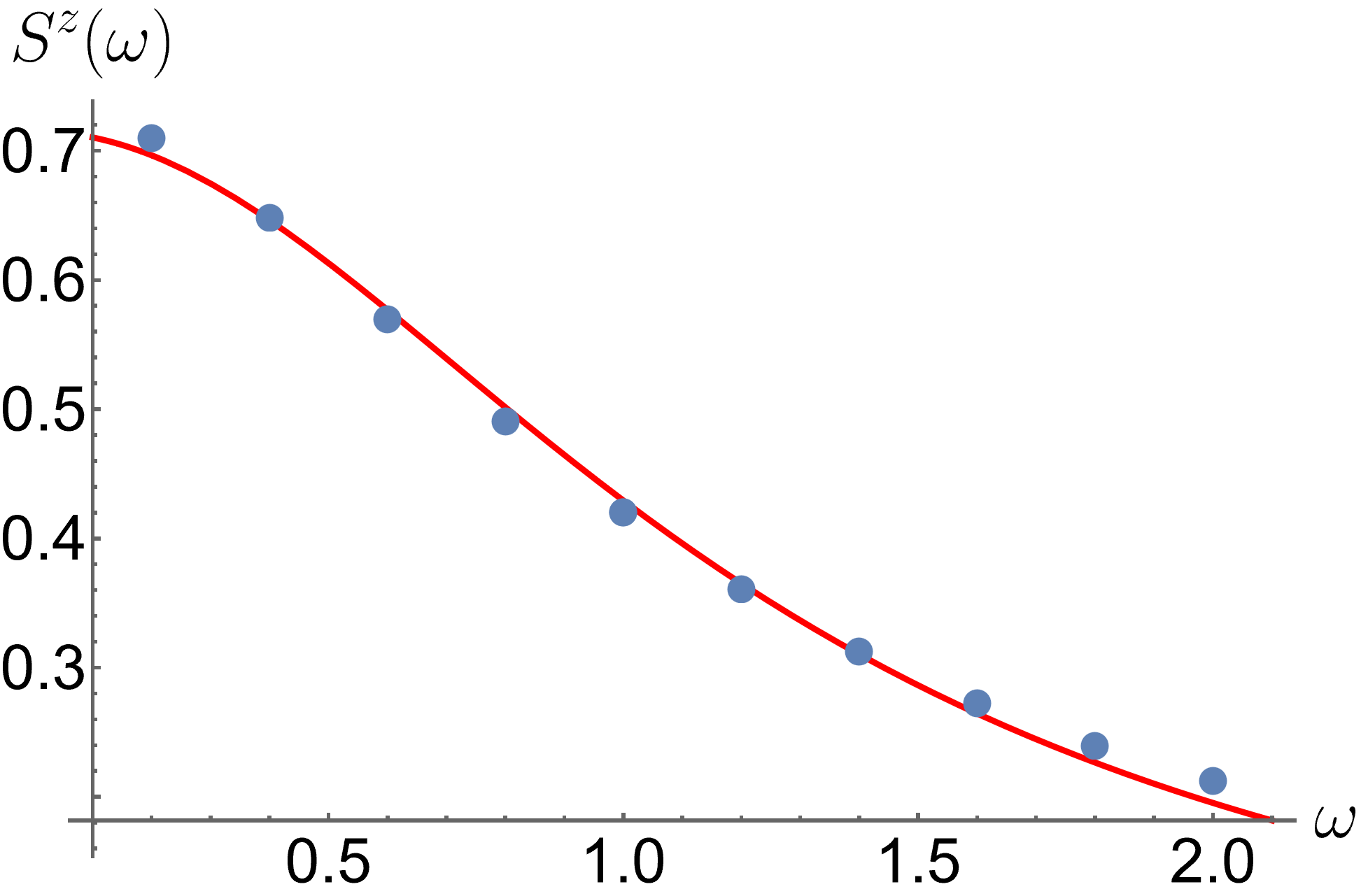}
\caption{The PSD \eqref{spectrum-example} for $\gamma = 1$, evaluated at $\omega \in \lbrace 0.2, 0.4, \ldots, 2 \rbrace$ (blue dots) and a Lorentzian fit to the data (red line). \label{fig-spectrum-fit}}
\end{figure}
\begin{figure}
\includegraphics[width=0.47\textwidth]{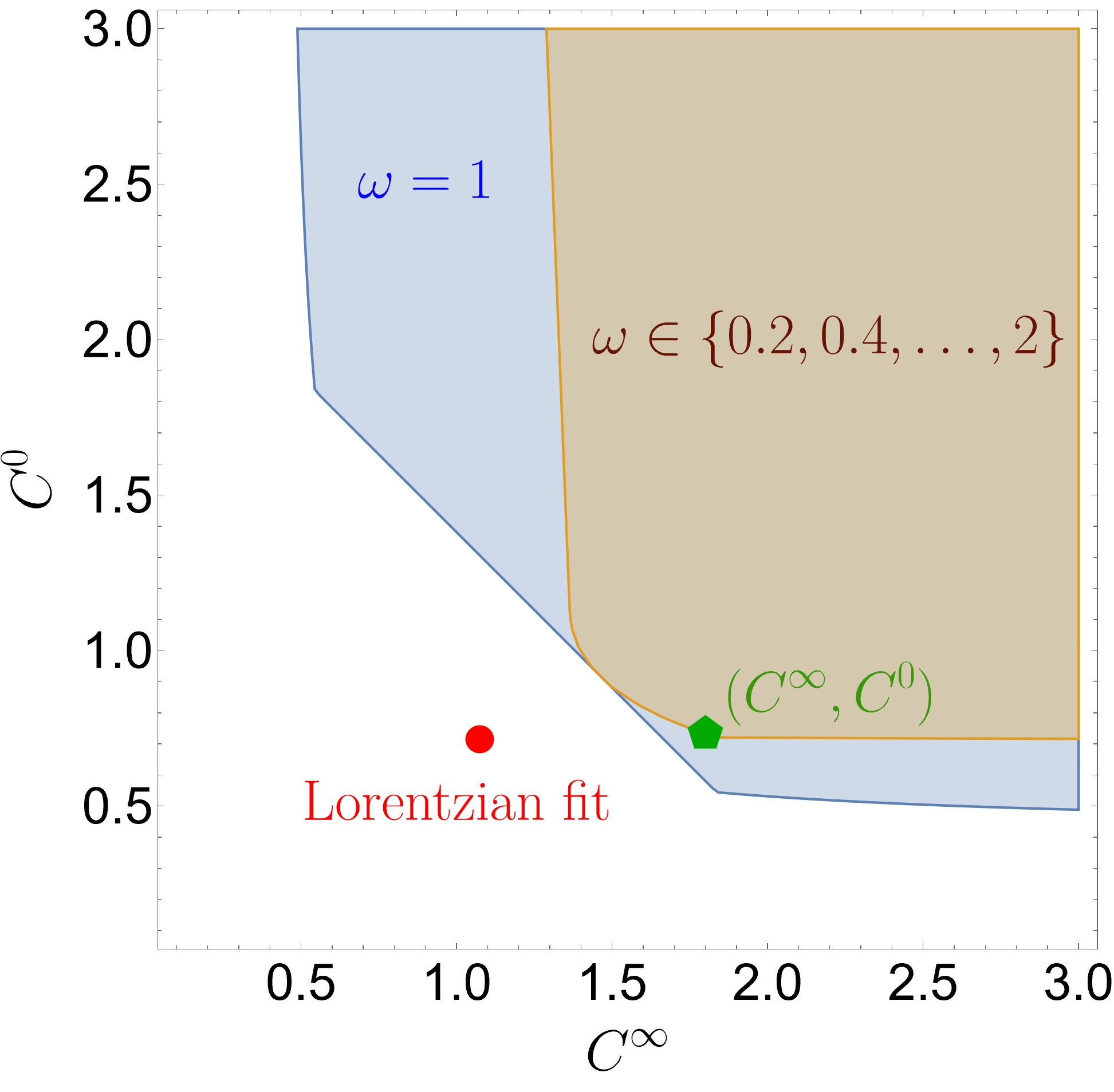}
\caption{The constraints on $C^0$ and $C^\infty$ obtained from \eqref{spectral-bound-lower-upper}, evaluated for the data shown in Fig.~\ref{fig-spectrum-fit}.
The blue shaded region corresponds to the values of $C^0$ and $C^\infty$ satisfying the inequality at $\omega = 1$, while the brown shaded region shows the values satisfying the inequality at all measured values of $\omega$. 
The red dot shows the parameters obtained from the Lorentzian fit, while the green pentagon corresponds to the actual asymptotic behavior of \eqref{spectrum-example}.
Surprisingly, the inequality at the single measurement point $\omega = 1$ is sufficient to verify that the Lorentzian fit cannot describe the spectrum of the system.} \label{fig-parameter-constraints}
\end{figure}

\textit{Constraints on asymptotic PSD.}
The lower and upper bound \eqref{spectral-bound-lower-upper} are not only useful to understand the relation between the PSD and dissipation, but can also be used to constrain the asymptotic behavior of the PSD.
In equilibrium, the bounds are expressed in terms of the low- and high-frequency asymptotes of the spectrum, see \eqref{c-equilibrium}, but this asymptotic behavior may not be accessible in practice, where typically only a finite range of frequencies can be observed.
As a concrete example, consider the PSD
\begin{align}
\frac{S^z(\omega)}{2 \text{Var}_\text{st}(z)} = \frac{\gamma (33 \gamma^2 + 9 \omega^2)}{45 \gamma^4 + 50 \gamma^2 \omega^2 + 5 \omega^4} \label{spectrum-example},
\end{align}
where $\gamma > 0$ is a quantity with dimensions of frequency.
This PSD is obtained for the observable $z(x) = x^3$ for a Brownian particle trapped in a parabolic potential $U(x) = \kappa x^2/2$, where $\gamma = \mu \kappa$ and $\mu$ is the particle mobility (see also Appendix S~IV).
Suppose that we measure the PSD for $\omega \in \lbrace 0.2, 0.4, \ldots, 2 \rbrace$, obtaining the data points shown in Fig.~\ref{fig-spectrum-fit}.
Since the PSD appears to consist of a single peak centered at zero frequency, we may attempt to fit it with a Lorentzian curve,
\begin{align}
\frac{\hat{S}^z(\omega)}{2 \text{Var}_\text{st}(z)} = \frac{1}{\frac{1}{\hat{C}^0} + \frac{\omega^2}{\hat{C}^\infty}} ,
\end{align}
which leads to reasonable agreement (red line in Fig.~\ref{fig-spectrum-fit}).
Here, we used the constants $\hat{C}^0$ and $\hat{C}^\infty$ as fit parameters, which also describe the extrapolated low- and high-frequency behavior of the PSD.
However, when using the obtained values in \eqref{spectral-bound-lower-upper}, we see that they do not satisfy the bounds.
This is illustrated in Fig.~\ref{fig-parameter-constraints}.
Consequently, the inequality constraints \eqref{spectral-bound-lower-upper} tell us that a single Lorentzian is not sufficient to describe the data, and, in particular, will not result a consistent extrapolation beyond the measured values.
This example demonstrates that, even in equilibrium, the bounds on the PSD can be used to verify whether the results obtained from a fit are compatible with the physical constraints on the system.

\textit{Discussion.}
We have derived bounds on the PSD, which extends the concept of constraints on the fluctuations of observables from the time to the frequency domain.
Surprisingly, we find that the same two constants provide both a lower and an upper bound on the PSD; in the time-domain, upper bounds on fluctuations have only been considered very recently \cite{Bak22}.
For equilibrium systems, these constants can be identified with the asymptotic low- and high-frequency behavior of the PSD, while for non-equilibrium systems they also contain information about the dissipation in the system.
This allows us to obtain estimates on the entropy production rate in terms of the measured spectrum via \eqref{spectral-tur-omega} and \eqref{spectral-tur-0}.

Recently, the eigenvalues of the generator of the dynamics have been related to thermodynamic quantities \cite{Obe22,Ohg23,Kol23}.
In particular, eigenvalues with a non-zero imaginary part can be used to infer the amount of entropy production \cite{Obe22,Kol23}.
Since the imaginary part also gives rise to oscillations, which appear as peaks in the PSD, a relation between \eqref{spectral-tur-omega} and these results appears reasonable.
We note, however, that generally, \eqref{spectral-tur-omega} can lead to a non-zero estimate on the entropy production rate even if all eigenvalues are real; conversely, even in the presence of complex eigenvalues, \eqref{spectral-tur-omega} may give a trivial result depending on the observable. 

Finally, we remark on possible extensions of our results. 
While here, we only considered steady-state dynamics, it is natural to describe systems that are driven by time-periodic forces in terms of their PSD. 
Recently, it has been found \cite{Lac22} that the corresponding PSD can be decomposed into a discrete part reflecting the driving and a continuous background. 
We speculate that the latter is subject to constraints similar to the ones derived here. 
Second, while overdamped dynamics can exhibit oscillations only out of equilibrium, in underdamped dynamics, oscillations can occur both in and out of equilibrium. 
Thus, it would be interesting to investigate what types of constraints on the PSD apply in this case and whether they can distinguish reversible from irreversible oscillations. 
Third, complex systems often exhibit a feature called 1/f noise \cite{Hoo81,Kes82}, where the PSD diverges at low frequencies, reflecting long-ranged temporal correlations \cite{Nie13,Lei15,Dec15b}. 
While such systems are beyond the current approach, characterizing them in terms of bounds on the spectrum might allow setting general constraints on the occurrence and properties of 1/f noise.

\begin{acknowledgments}
A.~D.~is supported by JSPS KAKENHI (Grant No. 19H05795, and 22K13974). The author thanks S.~i.~Sasa and J. Garnier-Brun for discussions and comments on improving the manuscript.
\end{acknowledgments}

\appendix

\onecolumngrid

\renewcommand{\theequation}{S\arabic{equation}}
\renewcommand{\thesection}{S~\Roman{section}}

\section{Derivation of the variational formula} \label{sec-variational}
In principle, the variational formula for the spectral density can be obtained from a variational expression for the cumulant generating function of a time-integrated function, similar to the technique used in Ref.~\cite{Dec23}.
We will investigate this approach in detail in an upcoming publication.
Here, we pursue a more direct approach.
According to the Wiener-Khinchin theorem, the spectral density is related to the steady-state covariance via a Fourier-transform,
\begin{align}
S^z(\omega) = \int_{-\infty}^{\infty} dt \ e^{i \omega t} \text{Cov}(z(t),z(0)) \label{wiener-khinchin} .
\end{align}
Since the covariance is a symmetric function of time, we can also write this as
\begin{align}
S^z(\omega) = 2 \int_0^\infty dt \ \cos(\omega t) \text{Cov}(z(t),z(0)) .
\end{align}
Expressing the covariance in terms of the conditional probability density $p_t(\bm{x} \vert \bm{y})$, this is equal to
\begin{align}
S^z(\omega) = 2 \int_0^\infty dt \int d\bm{x} \int d\bm{y} \ \cos(\omega t) z(x) z(y) \big( p_t(\bm{x} \vert \bm{y}) - p_\text{st}(\bm{x}) \big) p_\text{st}(\bm{y}) .
\end{align}
The time-evolution of $p_t(\bm{x} \vert \bm{y})$ is given by the Fokker-Planck equation
\begin{align}
\partial_t p_t(\bm{x} \vert \bm{y}) = \mathcal{L}(\bm{x}) p_t(\bm{x} \vert \bm{y}) \qquad \text{with} \qquad \mathcal{L}(\bm{x}) g(\bm{x}) = -\grad_x \Big( \big( \bm{f}(\bm{x}) - \bm{B} \grad_x \big) g(\bm{x}) \Big)  
\end{align}
with initial condition $p_t(\bm{x} \vert \bm{y}) = \delta(\bm{x} - \bm{y})$.
Since the steady-state is independent of time and in the kernel of the Fokker-Planck operator, $\mathcal{L}(\bm{x}) p_\text{st}(\bm{x})$, we can subtract the steady-state probability on both sides,
\begin{align}
\partial_t \big( p_t(\bm{x} \vert \bm{y}) - p_\text{st}(\bm{x}) \big) = \mathcal{L}(\bm{x}) \big( p_t(\bm{x} \vert \bm{y}) - p_\text{st}(\bm{x}) \big) \label{fpe-transition} .
\end{align}
Equivalently, we may express the time-evolution in terms of the backward Fokker-Planck equation \cite{Ris86},
\begin{align}
\partial_t \big( p_t(\bm{x} \vert \bm{y}) - p_\text{st}(\bm{x}) \big) = \mathcal{L}^\dagger(\bm{y}) \big( p_t(\bm{x} \vert \bm{y}) - p_\text{st}(\bm{x}) \big) \qquad \text{with} \qquad \mathcal{L}^\dagger(\bm{y}) g(\bm{y}) = \big( \bm{f}(\bm{y}) \grad_y + \grad_y \bm{B} \grad_y \big) g(\bm{y}) \label{fpe-transition-backward} .
\end{align}
We now define the conditional fluctuation of $z(\bm{x})$,
\begin{align}
\xi_t(\bm{y}) = \int d\bm{x} \ z(\bm{x}) \big( p_t(\bm{x} \vert \bm{y}) - p_\text{st}(\bm{x}) \big) = \Av{z}_{t \vert \bm{y}} - \Av{z}_\text{st} .
\end{align}
Multiplying \eqref{fpe-transition-backward} by $z(\bm{x})$ and integrating over $\bm{x}$, we then find
\begin{align}
\partial_t \xi_t(\bm{y}) = \mathcal{L}^\dagger(\bm{y}) \xi_t(\bm{y}) \label{xi-backward} ,
\end{align}
so that the adjoint Fokker-Planck operator also determines the time-evolution of the conditional fluctuation $\xi_t(\bm{y})$.
Next, we define
\begin{align}
\Xi_\omega(\bm{y}) = \int_0^\infty dt \ \cos(\omega t) \xi_t(\bm{y}),
\end{align}
that is, the Fourier-cosine transform of the conditional fluctuation.
Multiplying \eqref{xi-backward} by $\cos(\omega t)$ and integrating over time, we find
\begin{align}
\lim_{\tau \rightarrow \infty} \big(\cos(\omega \tau) \xi_\tau(\bm{y}) \big) - \xi_0(\bm{y}) + \omega \int_0^\tau dt \ \sin(\omega t) \xi_t(\bm{y}) = \mathcal{L}^\dagger(\bm{y}) \Xi_\omega(\bm{y}) .
\end{align}
We note that, by definition $\xi_0(\bm{y}) = z(\bm{y}) - \av{z}_\text{st}$, while, in the long-time limit, $\xi_t(\bm{y})$ vanishes since $\lim_{\tau \rightarrow \infty} p_\tau(\bm{x} \vert \bm{y}) = p_\text{st}(\bm{x})$.
Defining
\begin{align}
\tilde{\Xi}_\omega(\bm{y}) = \int_0^\infty dt \ \sin(\omega t) \xi_t(\bm{y}),
\end{align}
multiplying \eqref{xi-backward} by $\sin(\omega t)$ and integrating, we arrive at the set of coupled differential equations
\begin{subequations}
\begin{align}
\mathcal{L}^\dagger(\bm{y}) \Xi_\omega(\bm{y}) &= \omega \tilde{\Xi}_\omega(\bm{y}) - \big( z(\bm{y}) - \av{z}_\text{st} \big) \\
\mathcal{L}^\dagger(\bm{y}) \tilde{\Xi}_\omega(\bm{y}) &= - \omega \Xi_\omega(\bm{y}) .
\end{align}
\end{subequations}
In terms of $\Xi_\omega(\bm{y})$, the spectral density is given by
\begin{align}
S^z(\omega) = 2 \int d\bm{y} \ \Xi_\omega(\bm{y}) z(\bm{y}) p_\text{st}(\bm{y}) = 2 \int d\bm{y} \ \Xi_\omega(\bm{y}) \big(z(\bm{y}) - \av{z}_\text{st} \big) p_\text{st}(\bm{y}) = 2 \text{Cov}_\text{st}(\Xi_\omega, z) .
\end{align}
Here, we used that $\av{\Xi_\omega}_\text{st} = \Av{\tilde{\Xi}_\omega} = 0$, which follows from the definition.
We note that the adjoint Fokker-Planck operator can be written in terms of the steady-state local mean velocity as
\begin{align}
\mathcal{L}^\dagger(y) g(\bm{y}) =  \Big( \bm{\nu}_\text{st}(\bm{y}) \grad_y + \big[ \grad_y \ln p_\text{st}(\bm{y}) \big] \bm{B} \grad_y + \grad_y \bm{B} \grad_y \Big) g(\bm{y}),
\end{align}
where we adopt the convention that differential operators enclosed in brackets only act on terms inside the brackets, i.~e.~ $[ \grad_y g(\bm{y}) ] h(\bm{y}) = h(\bm{y}) \grad_y g(\bm{y})$.
This can be further rewritten as
\begin{align}
\mathcal{L}^\dagger(y) g(\bm{y}) = \frac{1}{p_\text{st}(\bm{y})} \grad_y \Big( \big( [\grad_y g(\bm{y})] \bm{B} + \bm{\nu}_\text{st}(\bm{y}) g(\bm{y}) \big) p_\text{st}(\bm{y}) \Big) .
\end{align}
Here, we used that $\grad_y \big(\bm{\nu}_\text{st}(\bm{y}) p_\text{st}(\bm{y}) \big) = 0$ by definition of the steady state.
Thus, the above equations can be written as
\begin{subequations}
\begin{align}
\grad_y \Big( \big( [\grad_y \Xi_\omega(\bm{y})] \bm{B} + \bm{\nu}_\text{st}(\bm{y}) \Xi_\omega(\bm{y}) \big) p_\text{st}(\bm{y}) \Big) &= \Big(\omega \tilde{\Xi}_\omega(\bm{y}) - \big(z(\bm{y}) - \av{z}_\text{st} \big) \Big) p_\text{st}(\bm{y}) \label{Xi-equations-1} \\
\grad_y \Big( \big( [\grad_y \tilde{\Xi}_\omega(\bm{y})] \bm{B} + \bm{\nu}_\text{st}(\bm{y}) \tilde{\Xi}_\omega(\bm{y}) \big) p_\text{st}(\bm{y}) \Big) &= - \omega \Xi_\omega(\bm{y}) p_\text{st}(\bm{y}) \label{Xi-equations-2} .
\end{align} \label{Xi-equations}%
\end{subequations}
We note the relation $\av{g \bm{\nu}_\text{st} \grad g} = \frac{1}{2} \av{\bm{\nu}_\text{st} \grad (g^2)}_\text{st} = 0$ for any function $g(\bm{y})$, which follows from the steady-state condition.
Multiplying \eqref{Xi-equations-2} by $\tilde{\Xi}_\omega(\bm{y})$ and integrating, we obtain the condition
\begin{align}
\omega \text{Cov}_\text{st}(\Xi_\omega,\tilde{\Xi}_\omega) = \Av{\grad \tilde{\Xi}_\omega \bm{B} \grad \tilde{\Xi}_\omega}_\text{st} .
\end{align}
Multiplying \eqref{Xi-equations-1} by $\Xi_\omega(\bm{y})$, integrating and using this condition, we then obtain the equivalent expression for the spectral density
\begin{align}
S^z(\omega) = 2 \Big( \Av{\grad \Xi_\omega \bm{B} \grad \Xi_\omega}_\text{st} + \Av{\grad \tilde{\Xi}_\omega \bm{B} \grad \tilde{\Xi}_\omega}_\text{st} \Big) \label{spectral-Xi} . 
\end{align}
We introduce the auxiliary functions $\chi_\omega(\bm{y})$, $\eta_\omega(\bm{y})$, $\tilde{\chi}_\omega(\bm{y})$ and $\tilde{\eta}_\omega(\bm{y})$, which satisfy the equations
\begin{subequations}
\begin{align}
\grad_y \Big( \big( [\grad_y \chi_\omega(\bm{y})] \bm{B} + \bm{\nu}_\text{st}(\bm{y}) \eta_\omega(\bm{y}) \big) p_\text{st}(\bm{y}) \Big) &= \Big(-\omega \tilde{\eta}_\omega(\bm{y}) - \big( z(\bm{y}) - \av{z}_\text{st} \big) \Big) p_\text{st}(\bm{y}) \\
\grad_y \Big( \big( [\grad_y \eta_\omega(\bm{y})] \bm{B} + \bm{\nu}_\text{st}(\bm{y}) \chi_\omega(\bm{y}) \big) p_\text{st}(\bm{y}) \Big) &= -\omega \tilde{\chi}_\omega(\bm{y}) p_\text{st}(\bm{y}) \\
\grad_y \Big( \big( [\grad_y \tilde{\chi}_\omega(\bm{y})] \bm{B} + \bm{\nu}_\text{st}(\bm{y}) \tilde{\eta}_\omega(\bm{y}) \big) p_\text{st}(\bm{y}) \Big) &=  \omega \eta_\omega(\bm{y}) p_\text{st}(\bm{y}) \\
\grad_y \Big( \big( [\grad_y \tilde{\eta}_\omega(\bm{y})] \bm{B} + \bm{\nu}_\text{st}(\bm{y}) \tilde{\chi}_\omega(\bm{y}) \big) p_\text{st}(\bm{y}) \Big) &=  \omega \chi_\omega(\bm{y}) p_\text{st}(\bm{y}) .
\end{align} \label{chi-eta-equations}%
\end{subequations}
Taking the sum of the first and second, respectively third and fourth, line, we find that this reproduces \eqref{Xi-equations}, provided that we identify $\Xi_\omega(\bm{y}) = \chi_\omega(\bm{y}) + \eta_\omega(\bm{y})$ and $-\tilde{\Xi}_\omega(\bm{y}) = \tilde{\chi}_\omega(\bm{y}) + \tilde{\eta}_\omega(\bm{y})$.
Moreover, we obtain the conditions
\begin{subequations}
\begin{align}
\Av{\grad \chi_\omega \bm{B} \grad \chi_\omega}_\text{st} &= \text{Cov}_\text{st}(\chi_\omega,z) + \text{Cov}_\text{st}(\chi_\omega,\bm{\nu}_\text{st} \grad \eta_\omega) + \omega \text{Cov}_\text{st}(\chi_\omega,\tilde{\eta}_\omega) \\
\Av{\grad \eta_\omega \bm{B} \grad \chi_\omega}_\text{st} &= \omega \text{Cov}_\text{st}(\chi_\omega,\tilde{\chi}_\omega) \\
\Av{\grad \eta_\omega \bm{B} \grad \eta_\omega}_\text{st} &= \text{Cov}_\text{st}(\eta_\omega,\bm{\nu}_\text{st} \grad \chi_\omega) + \omega \text{Cov}_\text{st}(\tilde{\chi}_\omega,\eta_\omega) \\
\Av{\grad \tilde{\chi}_\omega \bm{B} \grad \tilde{\chi}_\omega}_\text{st} &= \text{Cov}_\text{st}(\tilde{\chi}_\omega,\bm{\nu}_\text{st} \grad \tilde{\eta}_\omega) - \omega \text{Cov}_\text{st}(\tilde{\chi}_\omega,\eta_\omega) \\
\Av{\grad \tilde{\eta}_\omega \bm{B} \grad \tilde{\chi}_\omega}_\text{st} &= - \omega \text{Cov}(\chi_\omega,\tilde{\chi}_\omega) \\
\Av{\grad \tilde{\eta}_\omega \bm{B} \grad \tilde{\eta}_\omega}_\text{st} &= \text{Cov}_\text{st}(\tilde{\eta}_\omega, \bm{\nu}_\text{st} \grad \tilde{\chi}_\omega) - \omega \text{Cov}_\text{st}(\tilde{\eta}_\omega,\chi_\omega) .
\end{align} \label{chi-eta-relations}
\end{subequations}
Next, consider the convex optimization problem,
\begin{align}
\Psi &= \sup_{\chi,\tilde{\chi}} \inf_{\eta,\tilde{\eta}} \Big[ \psi[ \chi,\tilde{\chi},\eta,\tilde{\eta} ] \Big] \qquad \text{with}\\
 \Psi[ \chi,\tilde{\chi},\eta,\tilde{\eta} ] &= 2 \Big( \text{Cov}_\text{st}(z,\chi) + \text{Cov}_\text{st}(\chi,\bm{\nu}_\text{st} \grad \eta) + \text{Cov}_\text{st}(\tilde{\chi},\bm{\nu}_\text{st} \grad \tilde{\eta}) + \omega \big( \text{Cov}_\text{st}(\chi,\tilde{\eta}) - \text{Cov}_\text{st}(\tilde{\chi},\eta) \big) \Big) \nn
& \hspace{2cm} + \Av{\grad \eta \bm{B} \grad \eta}_\text{st} + \Av{\grad \tilde{\eta} \bm{B} \grad \tilde{\eta}}_\text{st} - \Av{\grad \chi \bm{B} \grad \chi}_\text{st} - \Av{\grad \tilde{\chi} \bm{B} \grad \tilde{\chi}}_\text{st} \n .
\end{align}
It is straightforward to verify that the Euler-Lagrange equations for this problem are precisely \eqref{chi-eta-equations}, that is, the solution of \eqref{chi-eta-equations} is the optimizer of $\Psi(\omega)$.
Moreover, using \eqref{chi-eta-relations}, we can write
\begin{align}
\Psi = \psi[ \chi_\omega,\tilde{\chi}_\omega,\eta_\omega,\tilde{\eta}_\omega ] &= \Av{\grad \eta_\omega \bm{B} \grad \eta_\omega}_\text{st} + \Av{\grad \tilde{\eta}_\omega \bm{B} \grad \tilde{\eta}_\omega}_\text{st} + \Av{\grad \chi_\omega \bm{B} \grad \chi_\omega}_\text{st} + \Av{\grad \tilde{\chi}_\omega \bm{B} \grad \tilde{\chi}_\omega}_\text{st} \\
&= \Av{\grad \big(\chi_\omega + \eta_\omega \big) \bm{B} \grad \big(\chi_\omega + \eta_\omega \big)}_\text{st} + \Av{\grad \big(\tilde{\chi}_\omega + \tilde{\eta}_\omega \big) \bm{B} \grad \big(\tilde{\chi}_\omega + \tilde{\eta}_\omega \big)}_\text{st} \nn
&= \Av{\grad \Xi_\omega \bm{B} \grad \Xi_\omega}_\text{st} + \Av{\grad \tilde{\Xi}_\omega \bm{B} \grad \tilde{\Xi}_\omega}_\text{st} \n .
\end{align}
Comparing this to \eqref{spectral-Xi}, we therefore obtain the variational expression for the spectral density,
\begin{align}
S^z(\omega) = 2 \sup_{\chi,\tilde{\chi}} \inf_{\eta,\tilde{\eta}} \Big[ \psi[ \chi,\tilde{\chi},\eta,\tilde{\eta} ] \Big] .
\end{align}
The result \eqref{spectral-variational} in the main text follows by introducing the complex functions $\hat{\chi}(\bm{y}) = \chi(\bm{y}) + i \tilde{\chi}(\bm{y})$ and $\hat{\eta}(\bm{y}) = \eta(\bm{y}) + i \tilde{\eta}(\bm{y})$,
in terms of which we can write the above expression as
\begin{align}
S^z(\omega) = 2 \sup_{\hat{\chi} \in \mathbb{C}} \inf_{\hat{\eta} \in \mathbb{C}} \Bigg[ 2 \Re \big( \text{Cov}_\text{st}(z + \bm{\nu}_\text{st} \grad \hat{\eta} - i \omega \hat{\eta}, \hat{\chi}^*) \big) + \Av{\grad \hat{\eta}^* \bm{B} \grad \hat{\eta}}_\text{st} - \Av{\grad \hat{\chi}^* \bm{B} \grad \hat{\chi}}_\text{st} \Bigg] \label{spectral-variational-app}.
\end{align}

\section{Derivation of the lower and upper bounds} \label{sec-bounds}
To obtain a lower bound on the spectral density, we set $\hat{\chi}(\bm{x}) = \alpha z(\bm{x})$ in \eqref{spectral-variational}, where $\alpha$ is a real constant,
\begin{align}
S^z(\omega) \geq 2 \inf_{\eta,\tilde{\eta}} \Bigg[ 2 \alpha \Big( \text{Var}_\text{st}(z) + \text{Cov}_\text{st}(z,\bm{\nu}_\text{st} \grad \eta) + \omega \text{Cov}_\text{st}(z,\tilde{\eta})  \Big) + \Av{\grad \eta \bm{B} \grad \eta}_\text{st} + \Av{\grad \tilde{\eta} \bm{B} \grad \tilde{\eta}}_\text{st} - \alpha^2 \Av{\grad z \bm{B} \grad z} \Bigg] .
\end{align}
We now rescale $\eta(\bm{x}) \rightarrow \beta \eta(\bm{x})$ and $\tilde{\eta}(\bm{x}) \rightarrow \tilde{\beta} \tilde{\eta}(\bm{x})$, where $\beta$ and $\tilde{\beta}$ are real constants.
Minimizing with respect to $\beta$ and $\tilde{\beta}$, we find
\begin{align}
S^z(\omega) \geq 2 \inf_{\eta,\tilde{\eta}} \Bigg[ 2 \alpha \text{Var}_\text{st}(z) - \alpha^2 \Bigg( \frac{\text{Cov}_\text{st}(z,\bm{\nu}_\text{st} \grad \eta)^2}{\Av{\grad \eta \bm{B} \grad \eta}_\text{st}} + \omega^2 \frac{\text{Cov}_\text{st}(z,\tilde{\eta})^2}{\Av{\grad \tilde{\eta} \bm{B} \grad \tilde{\eta}}_\text{st}} + \Av{\grad z \bm{B} \grad z} \Bigg) \Bigg].
\end{align}
Maximizing this expression with respect to $\alpha$, we obtain
\begin{align}
S^z(\omega) \geq 2 \inf_{\eta,\tilde{\eta}} \Bigg[ \frac{\text{Var}_\text{st}(z)^2}{\Av{\grad z \bm{B} \grad z} + \frac{\text{Cov}_\text{st}(z,\bm{\nu}_\text{st} \grad \eta)^2}{\Av{\grad \eta \bm{B} \grad \eta}_\text{st}} + \omega^2 \frac{\text{Cov}_\text{st}(z,\tilde{\eta})^2}{\Av{\grad \tilde{\eta} \bm{B} \grad \tilde{\eta}}_\text{st}}} \Bigg] .
\end{align}
Defining the constants
\begin{align}
C^\infty = \frac{1}{\text{Var}_\text{st}(z)} \sup_{\eta} \Bigg[\Av{\grad z \bm{B} \grad z} + \frac{\text{Cov}_\text{st}(z,\bm{\nu}_\text{st} \grad \eta)^2}{\Av{\grad \eta \bm{B} \grad \eta}_\text{st}} \Bigg] \qquad \text{and} \qquad
C^0 = \frac{1}{\text{Var}_\text{st}(z)} \sup_{\tilde{\eta}} \Bigg[ \frac{\text{Cov}_\text{st}(z,\tilde{\eta})^2}{\Av{\grad \tilde{\eta} \bm{B} \grad \tilde{\eta}}_\text{st}} \Bigg] \label{c-definition} ,
\end{align}
we obtain \eqref{spectral-lower} of the main text,
\begin{align}
\frac{S^z(\omega)}{2 \text{Var}_\text{st}(z)} \geq \frac{1}{C^\infty + \omega^2 C^0} .
\end{align}
On the other hand, an upper bound can be obtained by choosing $\hat{\eta}(\bm{x}) = \beta i z(\bm{x})$ with some real constant $\beta$,
\begin{align}
S^z(\omega) \leq 2 \sup_{\chi,\tilde{\chi}} \Bigg[ 2 \Big(\text{Cov}_\text{st}(z,\chi) + \beta \text{Cov}_\text{st}(\tilde{\chi},\bm{\nu}_\text{st} \grad z) + \beta \omega \text{Cov}_\text{st}(z,\chi) \Big)  - \Av{\grad \chi \bm{B} \grad \chi}_\text{st}  - \Av{\grad \tilde{\chi} \bm{B} \grad \tilde{\chi}}_\text{st} + \beta^2 \Av{\grad z \bm{B} \grad z}_\text{st} \Bigg] .
\end{align}
Rescaling $\chi(\bm{x}) \rightarrow \alpha \chi(\bm{x})$ and $\tilde{\chi}(\bm{x}) \rightarrow \tilde{\alpha} \tilde{\chi}(\bm{x})$ and maximizing with respect to $\alpha$ and $\tilde{\alpha}$ yields,
\begin{align}
S^z(\omega) \leq 2 \sup_{\chi,\tilde{\chi}} \Bigg[ (1+\beta \omega)^2 \frac{\text{Cov}_\text{st}(z,\chi)^2}{\Av{\grad \chi \bm{B} \grad \chi}_\text{st}} + \beta^2 \frac{\text{Cov}_\text{st}(\tilde{\chi},\bm{\nu}_\text{st} \grad z)^2}{\Av{\grad \tilde{\chi} \bm{B} \grad \tilde{\chi}}_\text{st}} + \beta^2 \Av{\grad z \bm{B} \grad z}_\text{st} \Bigg] .
\end{align}
Minimizing with respect to $\beta$, we obtain
\begin{align}
S^z(\omega) \leq 2 \sup_{\chi,\tilde{\chi}} \Bigg[ \frac{1}{\frac{\Av{\grad \chi \bm{B} \grad \chi}_\text{st}}{\text{Cov}_\text{st}(z,\chi)^2} + \frac{\omega^2}{\Av{\grad z \bm{B} \grad z} + \frac{\text{Cov}_\text{st}(\tilde{\chi}, \bm{\nu}_\text{st} \grad z)^2}{\Av{\grad \tilde{\chi} \bm{B} \grad \tilde{\chi}}_\text{st}}}} \Bigg] .
\end{align}
Using that
\begin{align}
\text{Cov}_\text{st}(\tilde{\chi}, \bm{\nu}_\text{st} \grad z) = - \text{Cov}_\text{st}(z, \bm{\nu}_\text{st} \grad \tilde{\chi}),
\end{align}
which follows from integrating by parts and using the steady-state condition $\grad_x \big(\bm{\nu}_\text{st}(\bm{x}) p_\text{st}(\bm{x}) \big) = 0$, we obtain \eqref{spectral-upper} of the main text,
\begin{align}
\frac{S^z(\omega)}{2 \text{Var}_\text{st}(z)} \leq \frac{1}{\frac{1}{C^0} + \frac{\omega^2}{C^\infty}} .
\end{align}

\section{Spectral density in equilibrium} \label{sec-equilibrium}
We now specialize \eqref{spectral-variational} to the case of an equilibrium system, that is $\bm{\nu}_\text{st}(\bm{x}) = 0$.
In terms of the real and imaginary parts of $\hat{\chi}(\bm{x})$ and $\hat{\eta}(\bm{x})$, we have
\begin{align}
S_\text{eq}^z(\omega) &= 2 \sup_{\chi,\tilde{\chi}} \inf_{\eta, \tilde{\eta}} \Bigg[ 2 \Big( \text{Cov}_\text{st}(z,\chi) + \omega \text{Cov}_\text{st}(\tilde{\eta},\chi) - \omega \text{Cov}_\text{st}(\eta,\tilde{\chi}) \Big) \\
&\hspace{3cm} + \Av{\grad \eta \bm{B} \grad \eta}_\text{st} + \Av{\grad \tilde{\eta} \bm{B} \grad \tilde{\eta}}_\text{st} - \Av{\grad \chi \bm{B} \grad \chi}_\text{st} - \Av{\grad \tilde{\chi} \bm{B} \grad \tilde{\chi}}_\text{st} \Bigg] . \n
\end{align}
We rescale $\tilde{\chi}(\bm{x}) \rightarrow \tilde{\alpha} \tilde{\chi}(\bm{x})$ and maximize with respect to $\alpha$,
\begin{align}
S_\text{eq}^z(\omega) &= 2 \sup_{\chi,\tilde{\chi}} \inf_{\eta, \tilde{\eta}} \Bigg[ 2 \Big( \text{Cov}_\text{st}(z,\chi) + \omega \text{Cov}_\text{st}(\tilde{\eta},\chi) \Big) + \frac{\text{Cov}_\text{st}(\eta,\tilde{\chi})^2}{\Av{\grad \tilde{\chi} \bm{B} \grad \tilde{\chi}}_\text{st}} + \Av{\grad \eta \bm{B} \grad \eta}_\text{st}  + \Av{\grad \tilde{\eta} \bm{B} \grad \tilde{\eta}}_\text{st} - \Av{\grad \chi \bm{B} \grad \chi}_\text{st} \Bigg] .
\end{align}
We note that both terms involving $\eta(\bm{x})$ are positive, and they attain their minimal value of $0$ by choosing $\eta(\bm{x}) = 1$,
\begin{align}
S_\text{eq}^z(\omega) &= 2 \sup_{\chi} \inf_{\tilde{\eta}} \Bigg[ 2 \Big( \text{Cov}_\text{st}(z,\chi) + \omega \text{Cov}_\text{st}(\tilde{\eta},\chi) \Big)  + \Av{\grad \tilde{\eta} \bm{B} \grad \tilde{\eta}}_\text{st} - \Av{\grad \chi \bm{B} \grad \chi}_\text{st} \Bigg] .
\end{align}
We now rescale $\chi(\bm{x}) \rightarrow \alpha \chi(\bm{x})$ and $\tilde{\eta}(\bm{x}) \rightarrow \tilde{\beta} \tilde{\eta}(\bm{x})$ and maximize/minimize with respect to $\alpha$ and $\beta$,
\begin{align}
S^z_\text{eq}(\omega) &= 2 \sup_{\chi} \inf_{\eta} \Bigg[ \frac{\text{Cov}(z,\chi)^2}{\Av{\grad \chi \bm{B} \grad \chi}_\text{st} + \omega^2 \frac{\text{Cov}_\text{st}(\eta,\chi)^2}{\Av{\grad \eta \bm{B} \grad \eta}_\text{st}}} \Bigg] = 2 \sup_{\chi} \Bigg[ \frac{\text{Cov}(z,\chi)^2}{\Av{\grad \chi \bm{B} \grad \chi}_\text{st} + \omega^2 \sup_{\eta} \Big[ \frac{\text{Cov}_\text{st}(\eta,\chi)^2}{\Av{\grad \eta \bm{B} \grad \eta}_\text{st}} \Big]} \Bigg] ,
\end{align}
where we renamed $\tilde{\eta}(\bm{x})$ to $\eta(\bm{x})$.
This is \eqref{spectral-eq-variational} of the main text.
Note that, even though the maximizer $\chi_\omega(\bm{x})$ generally depends on $\omega$ in a non-trivial manner, since both terms in the denominator are positive, it is clear that this is a decreasing function of $\omega$.
We can arrive at the same conclusion from the following argument:
In equilibrium, the Fokker-Planck operator is self-adjoint, so all its eigenvalues are real and (with the exception of the eigenvalue $0$ corresponding to the steady state) negative.
Moreover, we can find a set of orthogonal eigenfunctions.
Thus, the correlation function of any observable is a weighted sum of decaying exponential functions and thus itself monotonously decaying, see also Ref.~\cite{Tu08}.
Taking the Fourier transform, we obtain as the spectral density a weighted sum of Lorentizans centered at $\omega = 0$, each of which is a monotonously decaying function; thus, the spectral density itself is also a monotonously decaying function.

\section{Example: Brownian gyrator} \label{sec-parabolic}
As a concrete example, we consider the Brownian gyrator, which is an overdamped Brownian particle (mobility $\mu$ and temperature $T$), trapped in a two-dimensional parabolic potential $U(x_1,x_2) = \kappa (x_1^2 + x_2^2)$.
In addition, the particle is driven by the non-conservative force $\bm{F}^\text{nc}(x_1,x_2) = \epsilon (-x_2,x_1)$.
The corresponding Langevin equation is
\begin{subequations}
\begin{align}
\dot{x}_1(t) &= -\mu \kappa x_1(t) - \mu \epsilon x_2(t) + \sqrt{2 \mu T} \xi_1(t) \\
\dot{x}_2(t) &= -\mu \kappa x_2(t) + \mu \epsilon x_1(t) + \sqrt{2 \mu T} \xi_2(t) ,
\end{align}
\end{subequations}
where $\xi_1(t)$ and $\xi_2(t)$ are two independent Gaussian white noises.
The steady state of this system is
\begin{align}
p_\text{st}(x_1,x_2) = \frac{\kappa}{2\pi T} e^{-\frac{\kappa}{2 T}(x_1^2 + x_2^2)}, \qquad \bm{\nu}_\text{st}(x_1,x_2) = \mu \epsilon \begin{pmatrix} -x_2 \\ x_1 \end{pmatrix} . \label{steady-parabolic}
\end{align}
The system is out of equilibrium whenever $\epsilon \neq 0$, and the steady-state entropy production rate is given by
\begin{align}
\sigma_\text{st} = \frac{2 \mu \epsilon^2}{\kappa} .
\end{align}
As shown in Ref.~\cite{Dec23} we can further explicitly compute the transition probability density,
\begin{align}
p_t(x_1,x_2 \vert y_1,y_2) = \frac{\kappa}{2 \pi T ( 1 - e^{-2 \mu \kappa t}) } \exp \Bigg[ - \frac{\kappa}{2 T(1 - e^{-2 \mu \kappa t})} \bigg( \Big( &x_1 - e^{-\mu \kappa t} \big( y_1 \cos(\mu \epsilon t) - y_2 \sin(\mu \epsilon t) \big) \Big)^2 \label{transition-parabolic} \\
& + \Big( x_2 - e^{-\mu \kappa t} \big( y_2 \cos(\mu \epsilon t) + y_1 \sin(\mu \epsilon t) \big) \Big)^2 \bigg) \Bigg] . \n
\end{align}
Using \eqref{steady-parabolic} and \eqref{transition-parabolic}, we can compute arbitrary correlation functions, and, by using the Wiener-Khinchin theorem \eqref{wiener-khinchin}, the corresponding spectral density.
Here, we focus on the specific case of the observable $z(x_1,x_2) = x_1^3$.
The steady-state covariance is given by
\begin{align}
\text{Cov}(z(t),z(0) &= \int dx_1 \int dx_2 \int dy_1 \int dy_2 \ x_1^3 y_1^3 p_t(x_1,x_2 \vert y_1,y_2) p_\text{st}(y_1,y_2) \\
& = \frac{3 T^3}{\kappa^3} \Big( 3 e^{-\mu \kappa t} \cos(\mu \epsilon t) + 2 e^{-3 \kappa t} \cos(\mu \epsilon t)^3 \Big). \n
\end{align}
Taking the Fourier transform of this expression, we obtain 
\begin{align}
S^z(\omega) = \frac{6 T^3}{\kappa^3} \Bigg( &\frac{3 \gamma \big( \gamma^2 + \delta^2 + \omega^2 \big)}{\big(\gamma^2 + (\omega - \delta)^2\big) \big(\gamma^2 + (\omega + \delta)^2 \big)} + \frac{18 \gamma \big(9 \gamma^2 + \delta^2 + \omega^2 \big)}{\big(9 \gamma^2 + (\omega - \delta)^2\big) \big(9 \gamma^2 + (\omega + \delta)^2 \big)} \\
& + \frac{6 \gamma \big(9 \kappa^2 + 9 \delta^2 + \omega^2 \big)}{\big(9 \gamma^2 + (\omega - 3 \delta)^2\big) \big(9 \gamma^2 + (\omega + 3 \delta)^2 \big) } \Bigg) \n ,
\end{align}
where we defined $\gamma = \mu \kappa$ and $\delta = \mu \epsilon$.
Using the steady-state variance of $z(x_1,x_2)$,
\begin{align}
\text{Var}_\text{st}(z) = \frac{15 T^3}{\kappa^3},
\end{align}
we obtain for the normalized spectral density,
\begin{align}
\frac{S^z(\omega)}{2 \text{Var}_\text{st}(z)} =  \frac{1}{5} \Bigg( &\frac{3 \gamma \big( \gamma^2 + \delta^2 + \omega^2 \big)}{\big(\gamma^2 + (\omega - \delta)^2\big) \big(\gamma^2 + (\omega + \delta)^2 \big)} + \frac{18 \gamma \big(9 \gamma^2 + \delta^2 + \omega^2 \big)}{\big(9 \gamma^2 + (\omega - \delta)^2\big) \big(9 \gamma^2 + (\omega + \delta)^2 \big)} \label{spectrum-parabolic} \\
& + \frac{6 \gamma \big(9 \kappa^2 + 9 \delta^2 + \omega^2 \big)}{\big(9 \gamma^2 + (\omega - 3 \delta)^2\big) \big(9 \gamma^2 + (\omega + 3 \delta)^2 \big) } \Bigg) \n ,
\end{align}
In equilibrium ($\delta = 0$), this simplifies to
\begin{align}
\frac{S^z_\text{eq}(\omega)}{2 \text{Var}_\text{st}(z)} = \frac{\gamma (33 \gamma^2 + 9 \omega^2)}{45 \gamma^4 + 50 \gamma^2 \omega^2 + 5 \omega^4} \label{spectrum-parabolic-eq},
\end{align}
which is \eqref{spectrum-example} of the main text.
Next, we compute the corresponding values of $C^0$ and $C^\infty$.
As shown in the main text, we have
\begin{align}
C^0 &= C^0_\text{eq} = \frac{S_\text{eq}^z(0)}{2 \text{Var}_\text{st}(z)} = \frac{11}{15 \gamma} \label{c0-parabolic} . \\
C^\infty_\text{eq} &= \lim_{\omega \rightarrow \infty} \bigg( \frac{ \omega^2 S^z_\text{eq}(\omega)}{2 \text{Var}_\text{st}(z)} \bigg) = \frac{9}{5} \gamma  \label{ci-eq}.
\end{align}
This can be further bounded by the spectral gap \cite{Dec23},
\begin{align}
C^0 \leq \lambda^* = \frac{1}{\gamma} \label{c0-bound} .
\end{align}
On the other hand, from \eqref{c-definition}, we have
\begin{align}
C^\infty= \frac{\mu T \Av{\Vert\grad z\Vert^2 }}{\text{Var}_\text{st}(z)} + \sup_{\eta} \Bigg[ \frac{\text{Cov}_\text{st}(z,\bm{\nu}_\text{st} \grad \eta)^2}{\mu T \text{Var}_\text{st}(z) \Av{\Vert\grad \eta \Vert^2}_\text{st}} \Bigg] = \frac{9 \gamma}{5} + \sup_{\eta} \Bigg[ \frac{\text{Cov}_\text{st}(z,\bm{\nu}_\text{st} \grad \eta)^2}{\mu T \text{Var}_\text{st}(z) \Av{\Vert\grad \eta \Vert^2}_\text{st}} \Bigg] . 
\end{align}
To obtain an explicit bound for the second term, we write
\begin{align}
\frac{\text{Cov}_\text{st}(z,\bm{\nu}_\text{st} \grad \eta)^2}{\mu T \Av{\Vert\grad \eta \Vert^2}_\text{st}} = \frac{\Av{(z-z_0) \grad \eta \bm{\nu}_\text{st}}^2}{\mu T \Av{\Vert\grad \eta \Vert^2}_\text{st}} \leq \frac{1}{\mu T} \Av{(z-z_0)^2 \Vert \bm{\nu}_\text{st} \Vert^2}_\text{st},
\end{align}
where $z_0$ is an arbitrary constant and we used the Cauchy-Schwarz inequality.
Minimizing the expression on the right-hand side with respect to $z_0$, this can be written as
\begin{align}
\frac{\text{Cov}_\text{st}(z,\bm{\nu}_\text{st} \grad \eta)^2}{\mu T \Av{\Vert\grad \eta \Vert^2}_\text{st}} \leq \sigma_\text{st} \text{Var}_\sigma(z) ,
\end{align}
where $\text{Var}_\sigma$ denotes the variance with respect to the entropy-rescaled probability density,
\begin{align}
p_\sigma(x_1,x_2) = \frac{\Vert \bm{\nu}_\text{st}(x_1,x_2) \Vert^2}{\Av{\Vert \bm{\nu}_\text{st} \Vert^2}_\text{st}} p_\text{st}(x_1,x_2) .
\end{align}
This quantity can be evaluated using \eqref{steady-parabolic}, yielding
\begin{align}
\frac{\text{Var}_\sigma(z)}{\text{Var}_\text{st}(z)} = 8 .
\end{align}
We then have the upper bound,
\begin{align}
C^\infty \leq \frac{9 \gamma}{5} + \frac{8 \delta^2}{\gamma} \label{ci-bound} .
\end{align}
The spectrum shown in Fig.~\ref{fig-sd-bound} is obtained by using \eqref{spectrum-parabolic-eq}, \eqref{c0-parabolic} and \eqref{ci-eq} for the equilibrium case, and using \eqref{spectrum-parabolic}, \eqref{c0-bound} and \eqref{ci-bound} for the non-equilibrium case; the parameters are chosen as $\gamma = 1$ and $\delta = 3$.


%

\end{document}